\documentclass[12pt,aps,showpacs,eqsecnum,nofootinbib,floatfix]{revtex4}
\usepackage{CJK}
\usepackage{bbding}
\usepackage{pifont}
\usepackage{subfigure}
\usepackage{epsfig}
\usepackage{array}
\usepackage{amsmath}
\usepackage{amsfonts}
\usepackage{amssymb}
\usepackage{color}
\usepackage{graphicx}
\usepackage{mathrsfs}
\usepackage{multirow}
\usepackage{caption}

\def\be{\begin{equation}}
\def\ee{\end{equation}}
\def\bea{\begin{eqnarray}}
\def\eea{\end{eqnarray}}

\newcommand{\omits}[1]{}

\begin{document}

\title{The entropic force between two horizons of charged Gauss-Bonnet Black hole
in de Sitter Spacetime}
\author{Xiong-Ying Guo$^{a,b}$, Ying Gao$^{c}$, Huai-Fan Li$^{a,b}$\footnote{Email: huaifan.li@stu.xjtu.edu.cn; huaifan999@sxdtdx.edu.cn(H.-F. Li)}, Ren Zhao$^{b}$}

\medskip

\affiliation{\footnotesize$^a$ Department of Physics, Shanxi Datong
University,  Datong 037009, China\\
\footnotesize$^b$ Institute of Theoretical Physics, Shanxi Datong
University, Datong 037009, China\\
\footnotesize$^b$ School of Mathematics and Statistics, Shanxi Datong
University, Datong 037009, China}

\begin{abstract}
The basic equations of the thermodynamic system give the relationship between the internal energy, entropy and volume of two neighboring equilibrium states. By using the functional relationship between the state parameters in the basic equation, we give the differential equation satisfied by the entropy of spacetime. We can obtain the expression of the entropy by solving the differential equationy. This expression is the sum of entropy corresponding to the two event horizons and the interaction term. The interaction term is a function of the ratio of the locations of the black hole horizon and the cosmological horizon. The entropic force, which is strikingly similar to the Lennard-Jones force between particles, varies with the ratio of the two event horizons. The discovery of this phenomenon makes us realize that the entropic force between the two horizons may be one of the candidates to promote the expansion of the universe. 
\end{abstract}

 \maketitle

\section{Introduction}
 In the early period of inflation, our universe was in a quasi-de Sitter space. On the other hand, with the inclusion of mysterious component with negative pressure, a large number of dark energy models have been proposed to explain the cosmic acceleration. The simplest candidate for dark energy is the cosmological constant  (or vacuum energy density), with which our universe will naturally evolve into a new de Sitter phase.   Finally, there has also been flourishing interest in the duality relation of de Sitter space, promoted by the recent success of AdS/CFT correspondence in theoretical physics.  Therefore, from observational and theoretical point of view, it is rewarding to have a better understanding of the classical and quantum properties of de Sitter space~\cite{Cai,Mbarek,Simovic,Simovic2,Dolan,Sekiwa,Urano,Aiou,Kubiznak,McInerney,Bhattacharya,Zhang,Cai2,Pappas,Kanti,Bhattacharya2,Zhang2,Zhang3,Ma,Dinsmore,Chabab,Guo,Zhao}.

One of the most promising modified gravity theories is the Gauss--Bonnet 
(GB) gravity (also referred to as Einstein--GB gravity), which offers the 
leading order correction to the Einstein gravity. The GB term $\alpha$ is exactly 
the second order term in the Lagrangian of the most general Lovelock 
gravity. Therefore, although $\alpha$ itself is quadratic in curvature tensors, 
the equations of gravitational fields are still of second order and 
naturally avoid ghosts. The GB gravity possesses many important physical 
properties and has been heavily studied in gravitation and cosmology, also 
with emphasis in the extended phase space~\cite{Chen,Doneva,Hendi,Zhang20,Konoplya,Chen2,Cai3,Ali,Witek,Wei,Wang,Cai01,Kanti2,Doneva2,Wei2,Xu13,Zou,Miao18,Kleihaus}.

In this paper, based on the fact that de Sitter spacetime satisfies the first law of thermodynamics, the effective temperature and entropy of spacetime are obtained. Using the relationship between entropy and force~\cite{Verlind,Panos,Calderon,Komatsu,Komatsu2,Komatsu3,Caiy1,Caiy2}, we discussed that the entropic force between the black hole horizon and the cosmological horizon of charged Gauss-Bonnet black hole in de Sitter Spacetime(CGBDS). Since the entropy caused by the interaction between the two horizons is composed of two parts, the entropic force between the two horizons is also composed of two parts. Some of them are proportional to the GB factor. The results show that the entropic force between the two horizons in CGBDS is not only related to the ratio of the location of the two horizons, but also related to the GB terms.

By studying the entropic forces of the two parts of CGBDS with respect to the ratio of the locations of the event horizon, we find that they are strikingly similar to the Lennard-Jones force between particles with respect to the ratio of the coordinate locations of the two particles. When the two event horizons are close to each other, that is, when the spatial distance between the two event horizons is small, the cosmic horizon accelerates away from the black holes horizon under the action of entropic force. When the locations of the two horizons are relatively small, that is, the spatial distance between the two horizons is large, the separation speed of the two horizons slows down under the action of entropic force. The discovery of this phenomenon makes us realize that the entropy of interaction between the event horizon of black hole and the cosmic horizon, and the entropic force generated between the two event horizons, may be one of the alternatives to accelerate the expansion of the universe, that is, it may be a manifestation of dark energy. In particular, it should be noted that in our conclusion, the entropic force between the two horizons is proportional to the GB factor, and the size of GB factor directly affects the entropic force between the two horizons. If entropic force is one of the candidates to promote the expansion of the universe, then the value of GB factor directly affects the speed of the expansion of the universe.

This paper in organized as follows. In Sec. \ref{cabh}, we briefly introduce that the thermodynamic quantities corresponding to the black hole horizon and the cosmological horizon in CGBDS, and give the conditions that are satisfied when the black hole horizon and the cosmic horizon have the same radiation temperature. In Sec. \ref{eal}, based on the condition that CGBDS state parameter satisfies the first law of thermodynamics, we give that the expression of the effective thermodynamic quantity of CGBDS system, and obtain the equivalent temperature expression of CGBDS. Moreover, we find the differential equation of the entropy of CGBDS system, and obtain the interaction term of entropy of CGBDS system by solving the differential equation. In Sec. \ref{laten}, we discussed the entropic force between the two horizons in CGBDS by using the entropic force relationship, and obtained the entropy expression between the two horizons. The curves of the entropic force with respect to the ratio of the two horizons are compared with the Lennard-Jones force with respect to the ratio of the coordinates of the two particles. The Sec. \ref{con} is a discussion and summary. For simplicity, we adopt the units $\hbar=c=k_B=G=1$ in this paper.

\section{Charged Gauss-Bonnet Black hole in de Sitter Spacetime}
\label{cabh}

Higher derivative curvature terms occur in many occasions, such as in the 
semiclassically quantum gravity and in the effective low-energy action of 
superstring theories. Among the many theories of gravity with higher 
derivative curvature terms, due to the special features the Gauss-Bonnet 
gravity has attract much interest. The thermodynamic properties and 
phase structures of GB-AdS black hole have been briefly discussed in~\cite{Cai01}. 
In Refs.~\cite{Cai3,Wei2}, the critical phenomena and phase transition of the charged 
GB-AdS black hole have been studied extensively. In this paper, we study the 
thermal properties of charged GB-dS black hole after considering the 
connections between the black hole horizon and the cosmological horizon. 

The action of $d$-dimensional Einstein-Gauss-Bonnet-Maxwell theory with 
a bare cosmological constant $\Lambda $ reads
\begin{equation}
\label{eq1}
I = \frac{1}{16\pi }\int {d^dx\sqrt { - g} } \left[ {R - 2\Lambda + \alpha 
(R_{\mu \nu \gamma \delta } R^{\mu \nu \gamma \delta } - 4R_{\mu \nu } 
R^{\mu \nu } + R^2) - 4\pi F_{\mu \nu } F^{\mu \nu }} \right],
\end{equation}
where the GB coupling $\alpha $ has dimension [length]$^{2}$ and can be 
identified with the inverse string tension with positive value if the theory 
is incorporated in string theory, thus we shall consider only the case 
$\alpha > 0$. $F_{\mu \nu } $ is the Maxwell field strength defined as 
$F_{\mu \nu } = \partial _\mu A_\mu - \partial _\nu A_\nu $ with vector 
potential $A_\mu $. In addition, let us mention here that the GB term is a 
topological term in $d = 4$ dimensions and has no dynamics in this case. 
Therefore we will consider $d \ge 5$ in what follows. 

The action admits a static black hole solution with metric
\begin{equation}
\label{eq2}
ds^2 = - f(r)dt^2 + f^{ - 1}(r)dr^2 + r^2h_{ij} dx^idx^j,
\end{equation}
where $h_{ij} dx^idx^j$ represent the line of a $d - 2$-dimensional maximal 
symmetric Einstein space with constant curvature $(d - 2)(d - 3)k$ and 
volume $\Sigma _k $. Without loss of the generality, one may take $k = 1$, $0$ 
and $ - 1$, corresponding to the spherical, Ricci fiat and hyperbolic 
topology of the black hole horizon, respectively. The metric function $f(r)$ 
is given by~\cite{Xu,Wei3,Mam}
\begin{equation}
\label{eq3}
f(r) = k + \frac{r^2}{2\tilde {\alpha }}\left[ {1 - \sqrt {1 + \frac{64\pi 
\tilde {\alpha }M}{(d - 2)\Sigma _k r^{d - 1}} - \frac{2\tilde {\alpha 
}Q^2}{(d - 2)(d - 3)r^{2d - 4}} + \frac{8\tilde {\alpha }\Lambda }{(d - 1)(d 
- 2)}} } \right],
\end{equation}
where $\tilde {\alpha } = (d - 3)(d - 4)\alpha $, $M$ and $Q$ are the mass 
and charge of black hole respectively, and pressure $P$
\begin{equation}
\label{eq4}
P = - \frac{\Lambda }{8\pi } = -\frac{(d - 1)(d - 2)}{16\pi l^2}.
\end{equation}
Note that in order to have a well-defined vacuum solution with $M = Q = 0$, 
the effective Gauss-Bonnet coefficient $\tilde {\alpha }$ and pressure $P$ 
have to satisfy the following constraint
\begin{equation}
\label{eq5}
\frac{64\pi \tilde {\alpha }P}{(d - 1)(d - 2)} \le 1.
\end{equation}
When $d = 5$,  the location of the event horizon of the black hole $r_+ $ and the location of the cosmic event horizon $r_c $ satisfy the relation $f(r_{ + ,c} ) = 0$. The equations $f(r_ + ) = 0$ and $f(r_c ) = 0$ are rearranged to
\begin{equation}
\label{eq6}
M = \frac{3\Sigma _k r_ + ^2 }{16\pi }\left( {k + \frac{k^2\tilde {\alpha 
}}{r_ + ^2 }} \right) - \frac{\Sigma _k r_ + ^4 \Lambda }{32\pi } + 
\frac{\Sigma _k Q^2}{64\pi r_ + ^2 }
\end{equation}
\begin{equation}
\label{eq7}
M = \frac{3\Sigma _k r_c^2 }{16\pi }\left( {k + \frac{k^2\tilde {\alpha 
}}{r_c^2 }} \right) - \frac{\Sigma _k r_c^4 \Lambda }{32\pi } + \frac{\Sigma 
_k Q^2}{64\pi r_c^2 }.
\end{equation}
From Eqs. (\ref{eq6}) and (\ref{eq7}), we can obtain
\begin{equation}
\label{eq8}
M = \frac{3\Sigma _k kr_c^2 x^2}{16\pi (1 + x^2)} + \frac{3\Sigma _k 
k^2\tilde {\alpha }}{16\pi } + \frac{\Sigma _k Q^2}{64\pi r_c^2 (1\mbox{ + 
}x^2)x^2}\left( {1\mbox{ + }x^2 + x^4} \right),
\end{equation}
\begin{equation}
\label{eq9}
\Lambda = \frac{6}{r_c^2 (1 - x^4)}k(1 - x^2) - \frac{Q^2(1 - x^2)}{2r_c^6 
x^2(1 - x^4)},
\end{equation}
where $x =r_+/r_c$. From Eqs. (\ref{eq3}), (\ref{eq8}) and (\ref{eq9}), we can obtain
\[
f'(r_ + ) =\frac{2kr_+(1 - x^2)}{(r_+^2 + 2\tilde\alpha k)(1 + x^2)} - \frac{Q^2[(1 + x^2) - 2x^4]}{6r_+^3 (r_+^2+ 2\tilde {\alpha }k)(1 + x^2)} 
\]
\begin{equation}
\label{eq10}
 = \frac{2kr_c x(1 - x^2)}{(r_c^2 x^2 + 2\tilde {\alpha }k)(1 + 
 	x^2)} - \frac{Q^2[(1 + x^2) - 2x^4]}{6(r_c^2 x^2 + 2\tilde {\alpha }k)r_c^3 
 	x^3(1 + x^2)},
\end{equation}
\[
f'(r_c ) = - \frac{2kr_c (1 - x^2)}{(r_c^2 + 2\tilde {\alpha }k)(1 + x^2)} - 
\frac{Q^2[x^2(1 + x^2) - 2]}{6r_c^3 x^2(r_c^2 + 2\tilde {\alpha }k)(1 + 
x^2)}
\]
\begin{equation}
\label{eq11}
=-\frac{2kr_+ x(1 - x^2)}{(r_+^2 + 2\tilde \alpha kx^2)(1 + x^2)} - 
\frac{x^3 Q^2(x^2(1 + x^2)-2)}{6r_+^3 (r_+^2 + 2\tilde \alpha k x^2)(1 + x^2)}.
\end{equation}
Some thermodynamic quantities associated with the cosmological horizon are
\begin{equation}
\label{eq12}
T_c = - \frac{f'(r_c )}{4\pi }
\quad
S_c = \frac{\Sigma _k r_c^3 }{4}\left( {1 + \frac{6\tilde {\alpha }k}{r_c^2 
}} \right),
\quad
\Phi_c = \frac{\Sigma _k r_c^4 }{4}
\end{equation}
$T_c$, $S_c$ and $\Phi _c$ denote the Hawking temperature, the entropy and 
the charged potential. For the black hole horizon, associated thermodynamic quantities are 
\begin{equation}
\label{eq13}
T_ + = \frac{f'(r_ + )}{4\pi },
\quad
S_ + = \frac{\Sigma _k r_c^3 x^3}{4}\left( {1 + \frac{6\alpha _c k}{r_c^2 
x^2}} \right),
\quad
\Phi_ + = \frac{\Sigma _k r_c^4 x^4}{4}.
\end{equation}
From Eqs. (\ref{eq1}) and (\ref{eq11}), we found that the charge $Q$ of spacetime 
meets 
\[
Q^2 = \frac{12kr_c^4 (1 + x)(r_c^2 x - 2\tilde {\alpha }k)x^3}{[r_c^2 (1 + 
	x)(1 + x + 3x^2 + x^3 + x^4) + 2\tilde {\alpha }k(1 + x^3)]}
\]
\begin{equation}
\label{eq14}
= \frac{12kr_ + ^4 (1 + x)(r_ + ^2 - 2x\tilde {\alpha }k)}{[r_+^2 
	(1 + x)(1 + x + 3x^2 + x^3 + x^4) + 2x^2\tilde {\alpha }k(1 + x^3)]},
\end{equation}
the radiation temperature at the event horizon of the black hole is the same as that at the cosmic event horizon, 
\[
T = T_ + = T_c = \frac{kr_c (1\mbox{ + }x)^2(1 - x^2)}{2\pi [r_c^2 (1 + x)(1 
	+ x + 3x^2 + x^3 + x^4) + 2\tilde {\alpha }k(1 + x^3)]}
\]
\begin{equation}
\label{eq15}
= \frac{k r_+ x(1+x)^2(1 - x^2)}{2\pi (r_ + ^2 (1 + x)(1+x+3x^2 + x^3 + x^4) + 2x^2\tilde\alpha k(1 + x^3))}.
\end{equation}

\section{The effective thermodynamics quantites of  black hole}
\label{eal}
\begin{figure}[!htbp]
	\center{\includegraphics[width=7.5cm,keepaspectratio]{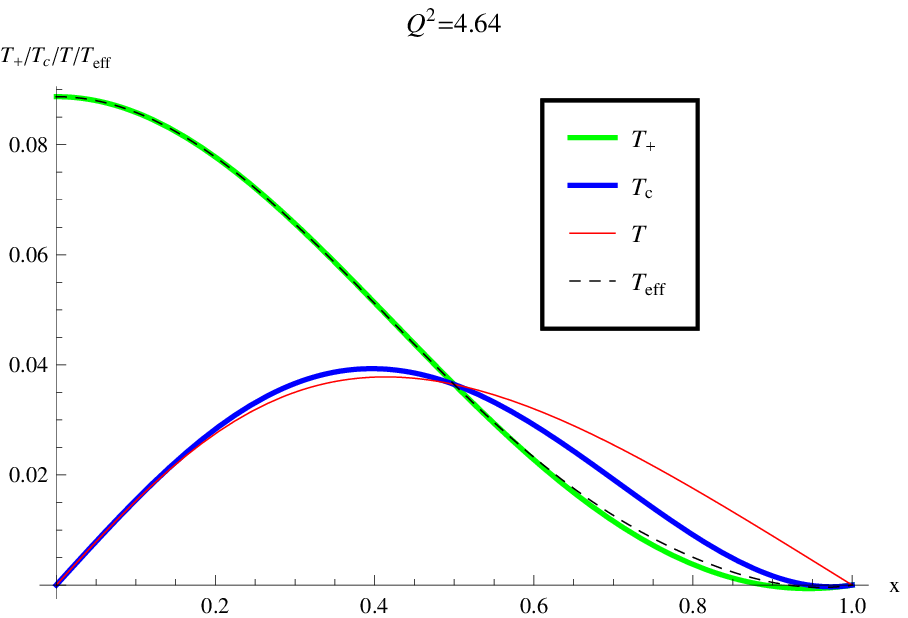}
		\includegraphics[width=7.5cm,keepaspectratio]{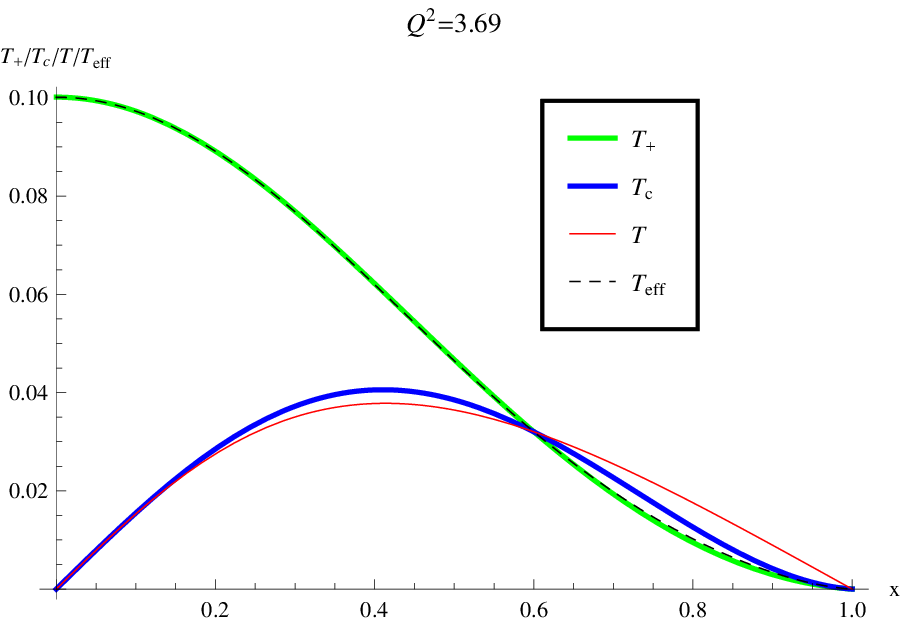}
		\includegraphics[width=7.5cm,keepaspectratio]{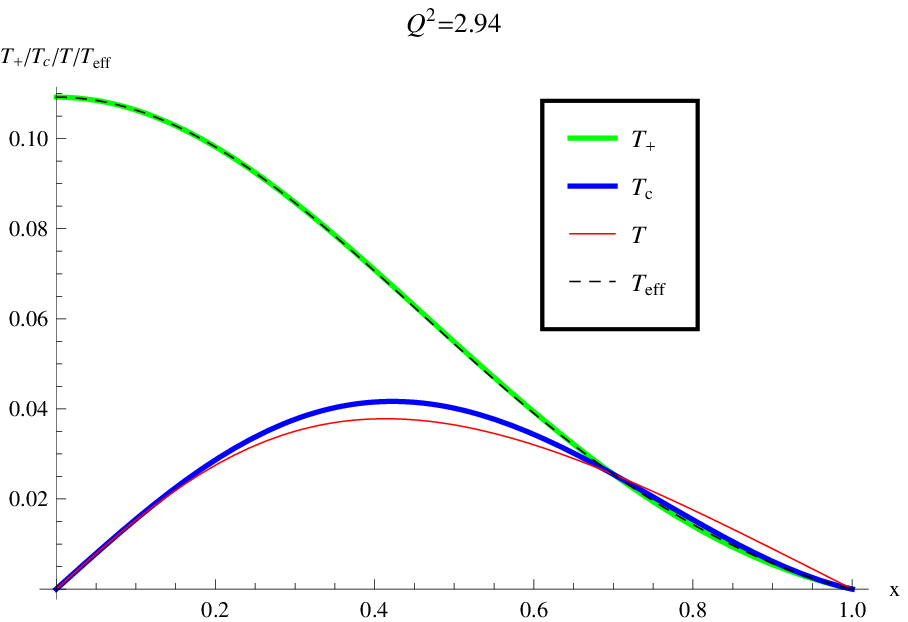}
		\includegraphics[width=7.5cm,keepaspectratio]{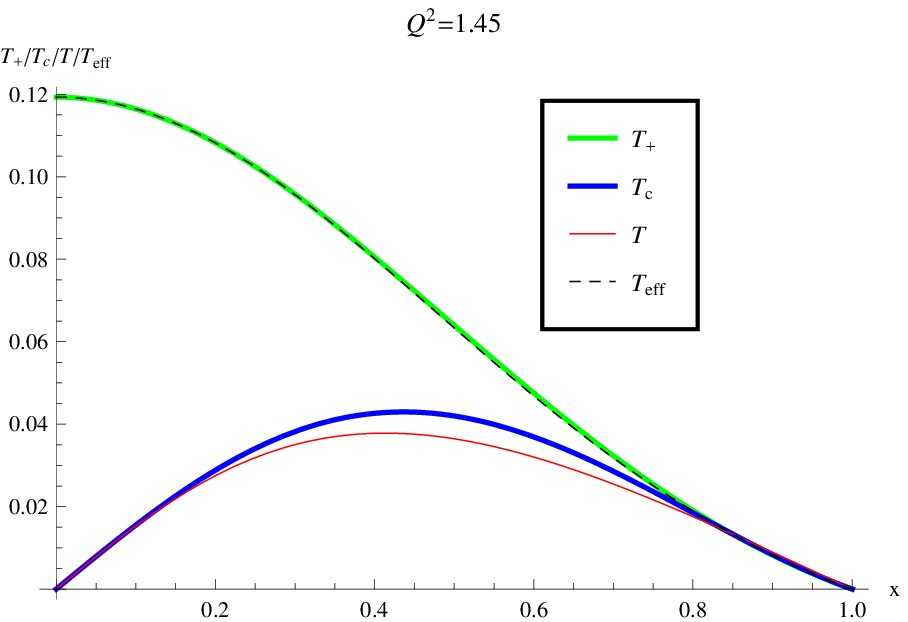}
		\hspace{0.5cm}}\\
	\captionsetup{font={scriptsize}}
	\caption{(Color online) The $T-x$ diagram for Charged Gauss-Bonnet black holes in de Sitter spacetime. \label{PVR}}
\end{figure}

\begin{figure}[!htbp]
\center{\includegraphics[width=7.5cm,keepaspectratio]{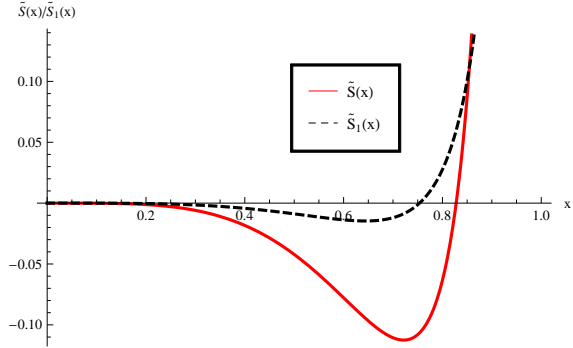}
\hspace{0.5cm}}\\
\captionsetup{font={scriptsize}}
\caption{(Color online) The entropy from the interaction between horizons of
	black holes and our universe. In the calcluate, we set $k\tilde \alpha=1$, $r_c=1$. \label{PVR1}}
\end{figure}

Considering the relation between the black hole horizon and the 
cosmological horizon, we can derive the effective thermodynamic quantities 
and corresponding first law of black hole thermodynamics
\begin{equation}
\label{eq16}
dM = T_{eff} dS - P_{eff} dV + \Phi _{eff} dQ,
\end{equation}
here the thermodynamic volume is that between the black hole horizon and the 
cosmological horizon, namely~\cite{Guo,Zhao,Dolan2}
\begin{equation}
\label{eq17}
V = V_c - V_ + = \frac{\Sigma _k }{4}r_c^4 \left( {1 - x^4} \right).
\end{equation}
Considering the expressions of entropy corresponding to the two horizons, as well as the dimensions and $\tilde {\alpha}k$ terms, we assume that the total entropy of spacetime is
\begin{equation}
\label{eq18}
S = \frac{\Sigma _k }{4}r_c^3 \left[ {f(x) + \frac{6\tilde {\alpha }k}{r_c^2 
}f_1 (x)} \right],
\end{equation}
here the function $f(x)$ and $f_1 (x)$ represents the extra 
contribution from the correlations of the two horizons. Taking $Q$, $\tilde{\alpha 
}$ as constant, substituting Eqs. (\ref{eq8}), (\ref{eq17}) and (\ref{eq18}) into Eq. (\ref{eq16}), we can obtain the effective temperature $T_{eff} $ of the system
\begin{equation}
\label{eq19}
T_{eff} = \frac{r_c 3kx(1 - x^2 + x^4) - \frac{Q^2}{4r_c^4 x^3}\left( 
{1 + x^2 - x^4 + x^6 + x^8} \right)}{2\pi (1 + x^2)\mbox{\{}r_c^2 
[f'(x)(1 - x^4) + 3x^3f(x)] + 6\tilde {\alpha }k[f_1 '(x)(1 - x^4) + x^3f_1 
(x)]\mbox{\}}}.
\end{equation}
When $Q^2$ satisfies equation (\ref{eq14}), the temperature corresponding to the two horizons is equal. In this case, we believe that the effective temperature of spacetime should also be radiation temperature, and
\begin{equation}
\label{eq20}
T_{eff} = \frac{kr_c (1\mbox{ + }x)^2(1 - x^2)}{2\pi [r_c^2 (1 + x)(1 + x + 
	3x^2 + x^3 + x^4) + 2\tilde {\alpha }k(1 + x^3)]}.
\end{equation}
From Eqs.(\ref{eq19}), (\ref{eq14}) and Eq. (\ref{eq20}), we can obtain
\[
\frac{\Sigma _k^2 r_c^9 k(1 - x^2)^2(1 + x^5)}{16\pi (1\mbox{ + 
}x^2)^2(r_c^2 x^2 + 2\tilde {\alpha }k)(r_c^2 + 2\tilde {\alpha }k)x}\mbox{ 
+ }\frac{\Sigma _k^2 r_c^7 k^2\tilde {\alpha }(1 - x^2)^2(1 + x^7)}{8\pi 
(1\mbox{ + }x^2)^2(r_c^2 x^2 + 2\tilde {\alpha }k)(r_c^2 + 2\tilde {\alpha 
}k)x^3}
\]
\[
= \frac{1}{4\pi r_c }\frac{kr_c^4 (1 - x^2)^2(1 - x^4)}{3x^3(1 + 
x^2)^2(r_c^2 + 2\tilde {\alpha }k)(r_c^2 x^2 + 2\tilde {\alpha }k)}\times 
\]
\begin{equation}
\label{eq21}
\left[ {\frac{\Sigma _k^2 }{4}r_c^6 [f'(x)(1 - x^4) + 3x^3f(x)] + 
\frac{3\Sigma _k^2 }{2}r_c^4 \tilde {\alpha }k[f_1 '(x)(1 - x^4) + x^3f_1 
(x)]} \right].
\end{equation}
Since $\tilde {\alpha }k$ in Eq. (\ref{eq21}) is an independent variable, the same terms at both sides of Eq. (\ref{eq21}) should be equal, i.e
\[
(1 - x^4)f'(x) + 3x^3f(x) = \frac{3x^2(1 + x^5)}{(1 - x^4)},
\]
\begin{equation}
\label{eq22}
(1 - x^4)f_1 '(x) + x^3f_1 (x) = \frac{(1 + x^7)}{(1 - x^4)}.
\end{equation}
Substituting Eq. (\ref{eq22}) into Eq. (\ref{eq19}) to obtain the effective temperature of spacetime
\[
T_{eff} = \frac{r_c kx(1 - 2x^2 + 2x^4 - x^6) - \frac{Q^2}{12r_c^3 x^3}(1 - 
	2x^4 + 2x^6 - x^{10})}{2\pi \left[ {r_c^2 x^2(1 + x^5) + 2\tilde {\alpha 
		}k(1 + x^7)} \right]}
\]
\begin{equation}
\label{eq23}
= \frac{r_ + k(1 - 2x^2 + 2x^4 - x^6) - Q^2(1 - 2x^4 + 2x^6 - x^{10}) / 
	(12r_ + ^3 )}{2\pi \left[ {r_ + ^2 (1 + x^5) + 2\tilde {\alpha }k(1 + x^7)} 
	\right]}.
\end{equation}
When $k =1$, $r_+ = 1$, $\tilde \alpha k =0.05$ , from Eqs.(\ref{eq12}), (\ref{eq13}), (\ref{eq15}) and (\ref{eq23}), we can plot the curve $T_{ + ,c}-x$, $T-x$ and $T_{eff}-x$ in Fig. \ref{PVR}.
From Fig.1, we can see that the intersection of curves $T_{+,c}-x$,  $T-x$ and $T_{eff}-x$ increases with the decrease of $Q^2$, and the intersection of the curve approaches $x\rightarrow1$.
When the initial conditions satisfy $f(0) = 1$, $f_1 (0) = 1$, the solution of Eq. (\ref{eq22}) is
\[
f(x) = \frac{11}{7}(1 - x^4)^{3 / 4} - \frac{4(1 + x^7) - 7x^3(1 + x)}{7(1 - 
x^4)}
\]
\begin{equation}
\label{eq24}
= \frac{11}{7}(1 - x^4)^{3 / 4} - \frac{4(1 + x^7)\mbox{ + }7(1 - 2x^4 - 
x^7)}{7(1 - x^4)}\mbox{ + 1 + }x^3 = \tilde {f}(x)\mbox{ + 1 + }x^3,
\end{equation}
\[
f_1 (x) = \frac{9(1 - x^4)^{1 / 4}}{5} - \frac{(4 - 5x - 5x^4 + 4x^5)}{5(1 - 
x^4)}
\]
\begin{equation}
\label{eq25}
= \frac{9(1 - x^4)^{1 / 4}}{5} - \frac{(9 - 10x^4 - x^5)}{5(1 - x^4)} + 1 + 
x = \tilde {f}_1 (x) + 1 + x.
\end{equation}
From Eqs.(\ref{eq12}), (\ref{eq13}), (\ref{eq18}), (\ref{eq24}) and (\ref{eq25}), we can obtain that the entropy of spacetime includes not only the sum of entropy $S_+ + S_c $ corresponding to the event horizon of the black hole and the cosmic event horizon, but also the entropy $\tilde {S} + \tilde {S}_1 $ caused by the interaction of the two event horizons
\begin{equation}
\label{eq26}
\tilde {S}\mbox{ = }\frac{\Sigma _k }{4}r_c^3 \tilde {f}(x),
\quad
\tilde {S}_1 = \frac{3\Sigma _k }{2}\tilde {\alpha }kr_c \tilde {f}_1 (x).
\end{equation}
From Eq.(\ref{eq26}), we can plot $\tilde {f}(x) - x$, $\tilde {f}_1 (x) - x$, $0 < x \le 1$ in Fig.\ref{PVR1}.
As can be seen from Fig.\ref{PVR1}, the two curves have the same change rule, and the amplitude of $\tilde S_1$ curve is proportional to $\tilde \alpha$.

From Eq.(\ref{eq16}), we can obtain the effective pressure $P_{eff} $ and the effective potential $\Phi_{eff}$
\[
P_{eff} = - \left( {\frac{\partial M}{\partial V}} \right)_{Q,S} = - 
\frac{\left( {\frac{\partial M}{\partial x}} \right)_{r_c } \left( 
{\frac{\partial S}{\partial r_c }} \right)_x - \left( {\frac{\partial 
M}{\partial r_c }} \right)_x \left( {\frac{\partial S}{\partial x}} 
\right)_{r_c } }{\left( {\frac{\partial V}{\partial x}} \right)_{r_c } 
\left( {\frac{\partial S}{\partial r_c }} \right)_x - \left( {\frac{\partial 
V}{\partial r_c }} \right)_x \left( {\frac{\partial S}{\partial x}} 
\right)_{r_c } }
\]
\[
= \frac{(1 - x^2)\left[ {\left[ {f(x)} \right] + 6\tilde {\alpha }kx^2 / r_ 
+ ^2 f_1 (x)} \right]}{8\pi (1 + x^2)\left[ {r_ + ^2 (1 + x^5) + 2\tilde 
{\alpha }k(1 + x^7)} \right]}\left[ {3kx - \frac{Q^2x(1 + 2x^2)}{4r_ + ^4 }} 
\right]
\]
\begin{equation}
\label{eq27}
- \frac{(1 - x^2)\left[ {f'(x) + 6\tilde {\alpha }k / r_c^2 f_1 '(x)} 
\right]}{24\pi \left[ {r_ + ^2 (1 + x^5) + 2\tilde {\alpha }k(1 + x^7)} 
\right]}\left[ {3kx^2 - \frac{Q^2x^2}{4r_ + ^4 }\left( {1\mbox{ + }x^2 + 
x^4} \right)} \right],
\end{equation}
\begin{equation}
\label{eq28}
\Phi _{eff} = \left( {\frac{\partial M}{\partial Q}} \right)_{S,V} = 
\frac{\Sigma _k Q}{32\pi r_ +^2 (1 +x^2)}\left( {1 + 
x^2 + x^4} \right).
\end{equation}

\section{the entropic force of two horizon of GBDST}
\label{laten}
The entropic force of the thermodynamic system is expressed as~\cite{Verlind,Komatsu,Komatsu2,Komatsu3,Caiy1,Caiy2}
\begin{equation}
\label{eq29}
F = - T\frac{\partial S}{\partial r},
\end{equation}
where $T$ is the temperature of system,  $r = r_c - r_ + = r_c (1 - x)$.
We consider the entropic force between the event horizon of the black hole and the event horizon of the universe. From Eq.(\ref{eq26}), we know that
the entropy caused by the interaction between the event horizon of the black hole and the cosmic event horizon is
\begin{equation}
\label{eq30}
\tilde {S}\mbox{ = }\frac{\Sigma _k }{4}r_c^3 \tilde {f}(x),
\quad
\tilde {S}_1 = \frac{3\Sigma _k }{2}\tilde {\alpha }kr_c \tilde {f}_1 (x).
\end{equation}
According to the entropic force relation Eq.(\ref{eq29}), we can obtain the entropic force of interaction between the two horizons can be expressed as
\begin{equation}
\label{eq31}
F=-T_{eff} \left( {\frac{\partial (\tilde {S} + \tilde {S}_1 )}{\partial 
		r}} \right)_{T_{eff} } ,
\end{equation}
where $T_{eff} $ is the effective temperature of the system. From Eq.(\ref{eq31}), we can obtain
\begin{equation}
\label{eq32}
F(x) = T_{eff} \frac{\left( {\frac{\partial (\tilde {S} + \tilde {S}_1 
)}{\partial r_c }} \right)_x \left( {\frac{\partial T_{eff} }{\partial x}} 
\right)_{r_c } - \left( {\frac{\partial (\tilde {S} + \tilde {S}_1 
)}{\partial x}} \right)_{r_c } \left( {\frac{\partial T_{eff} }{\partial r_c 
}} \right)_x }{(1 - x)\left( {\frac{\partial T_{eff} }{\partial x}} 
\right)_{r_c } + r_c \left( {\frac{\partial T_{eff} }{\partial r_c }} 
\right)_x } = \tilde {F}(x) + \tilde {F}_1 (x),
\end{equation}
where
\[
\tilde {F}(x) = T_{eff} \frac{\left( {\frac{\partial \tilde {S}}{\partial 
r_c }} \right)_x \left( {\frac{\partial T_{eff} }{\partial x}} \right)_{r_c 
} - \left( {\frac{\partial \tilde {S}}{\partial x}} \right)_{r_c } \left( 
{\frac{\partial T_{eff} }{\partial r_c }} \right)_x }{(1 - x)\left( 
{\frac{\partial T_{eff} }{\partial x}} \right)_{r_c } + r_c \left( 
{\frac{\partial T_{eff} }{\partial r_c }} \right)_x },
\]
\begin{equation}
\label{eq33}
\tilde {F}_1 (x) = T_{eff} \frac{\left( {\frac{\partial \tilde {S}_1 
}{\partial r_c }} \right)_x \left( {\frac{\partial T_{eff} }{\partial x}} 
\right)_{r_c } - \left( {\frac{\partial \tilde {S}_1 }{\partial x}} 
\right)_{r_c } \left( {\frac{\partial T_{eff} }{\partial r_c }} \right)_x 
}{(1 - x)\left( {\frac{\partial T_{eff} }{\partial x}} \right)_{r_c } + r_c 
\left( {\frac{\partial T_{eff} }{\partial r_c }} \right)_x }.
\end{equation}
The interaction between the two horizons is divided into two parts, $\tilde 
{F}(x)$ and $\tilde {F}_1 (x)$,  where $\tilde{F}(x)$ is caused by $\tilde S$, and $\tilde {F}_1 (x)$ is caused by  $\tilde S_1$. Since $\tilde S_1$ is proportional to GB factor $\tilde \alpha$, $\tilde {F}_1 (x)$ is proportional to $\tilde \alpha$. $\tilde {F}_1 (x)$ is greatly affected by GB factor $\tilde \alpha$, when $\tilde \alpha \rightarrow 0$, $\tilde {F}_1 (x)\rightarrow0$.

In order to more clearly reflect the change rule of entropic force with respect to the ratio $x$ between the two horizons, and the influence of different parameters on the entropic force ${F}(x)$ between the two horizons, so we can plot $\tilde{F}(x) - x$ and $\tilde {F}_1 (x) - x$ for different parameters.
\begin{figure}[!htbp]
\center{\includegraphics[width=5cm,keepaspectratio]{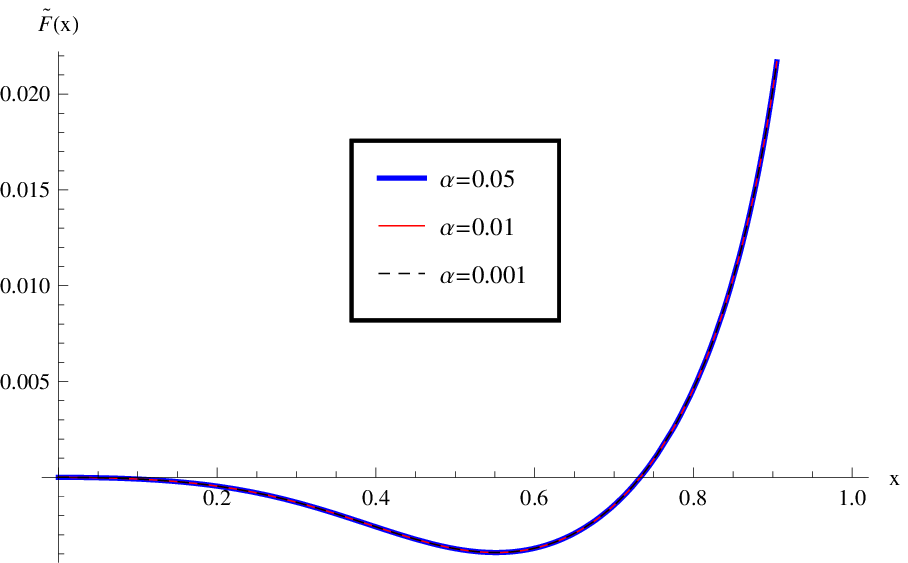}
\includegraphics[width=5cm,keepaspectratio]{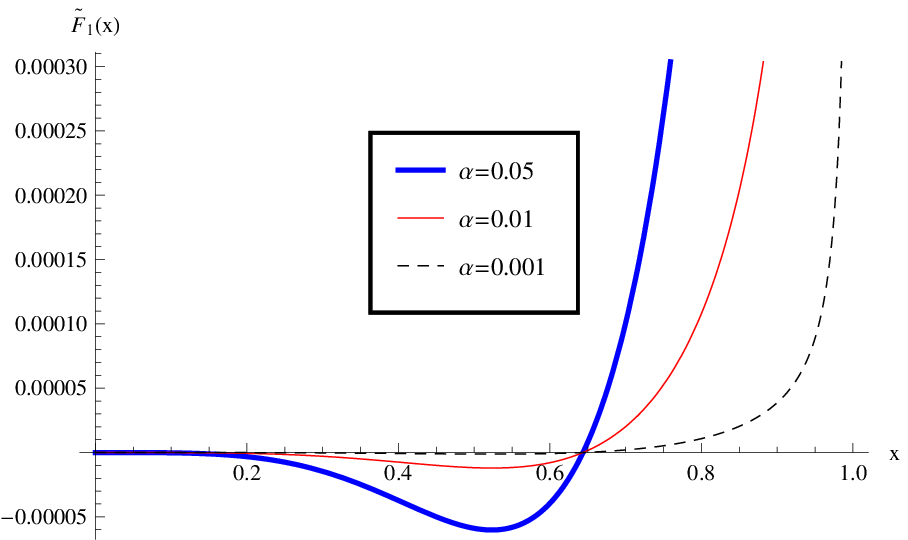}
\includegraphics[width=5cm,keepaspectratio]{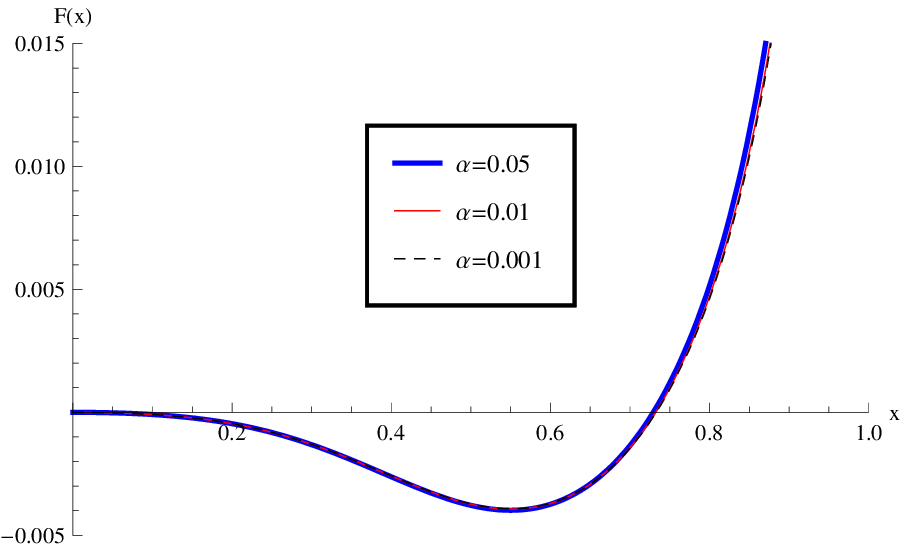}
\includegraphics[width=5cm,keepaspectratio]{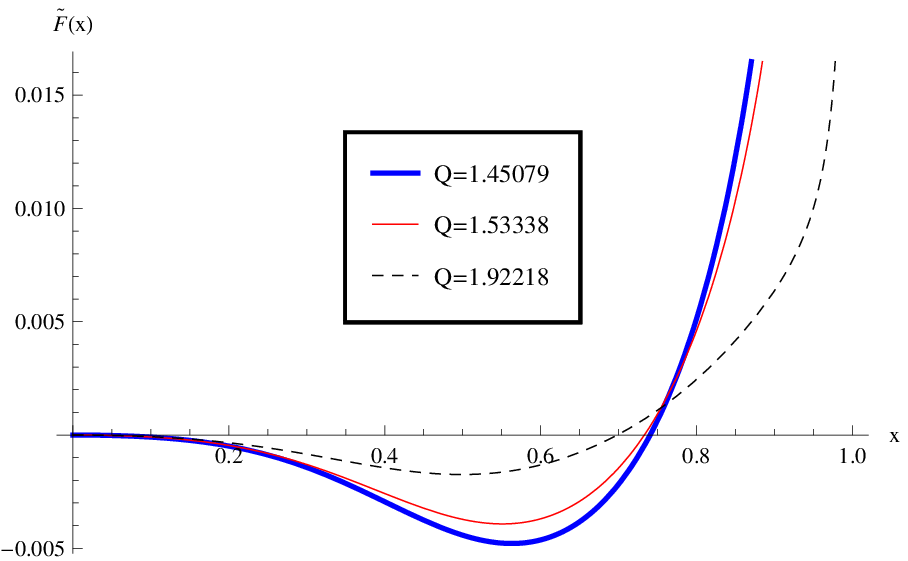}
\includegraphics[width=5cm,keepaspectratio]{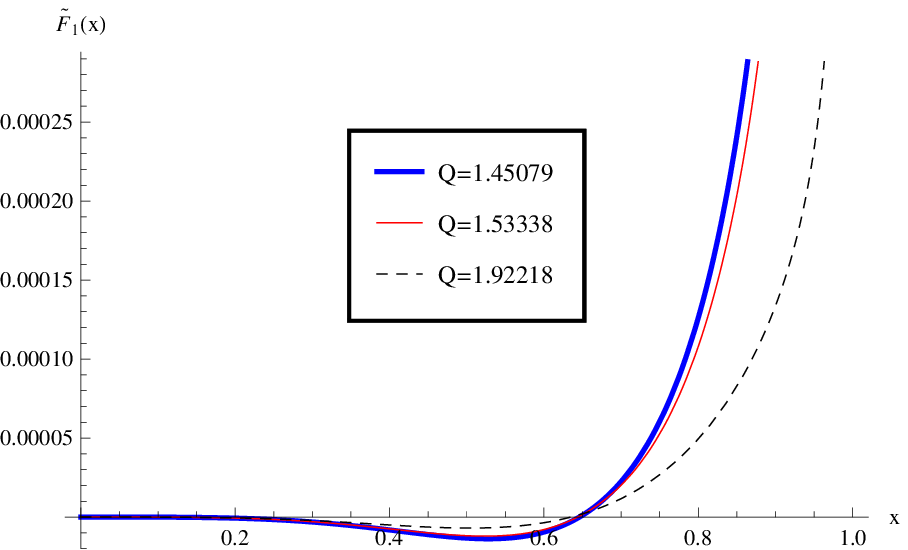}
\includegraphics[width=5cm,keepaspectratio]{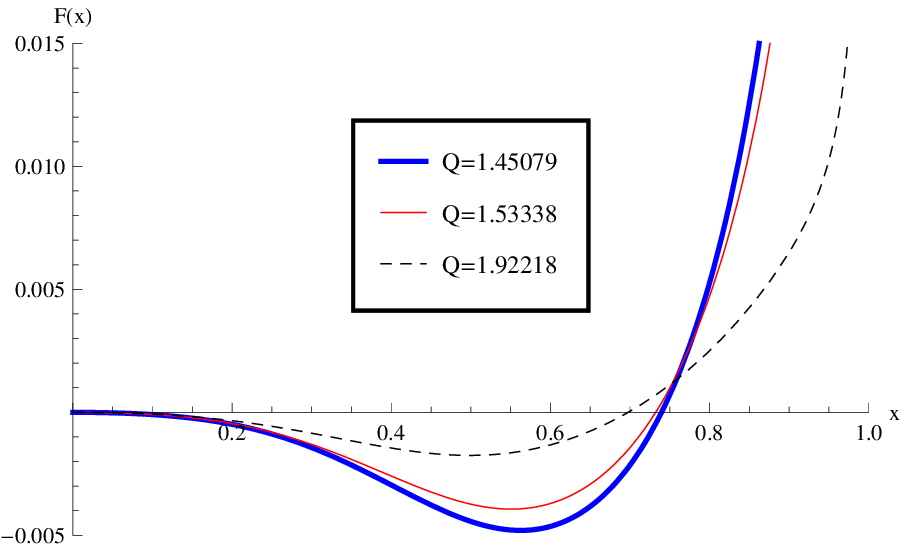}
\hspace{0.5cm}}\\
\captionsetup{font={scriptsize}}
\caption{(Color online) The entropic force changes with the ratio of the radius of the event horizon of the black hole to the radius of the cosmic event horizon for charged Gauss-Bonnet Black hole in de Sitter Spacetime with different Gauss-Bonnet factor $\tilde \alpha$ and charged $Q$.
\label{PVR3}}
\end{figure}
From Fig. \ref{PVR3}, we find that the variation curves of $\tilde{F}(x) - x$ and $\tilde {F}_1 (x) - x$ with respect to the location ratio of the two horizons are similar and have the same variation rules, and the amplitude of the curves is proportional to $\tilde \alpha$.

From Fig.\ref{PVR3}, the general change of entropic force with respect to $x$ is that as $x\rightarrow1$, the entropic force goes to infinity.  It is shown that when the event horizon of the black hole in de Sitter is close to the cosmic event horizon, the two event horizons are affected by the infinite entropic force, which accelerates the separation between the two event horizons. This corresponds to what we now think of as the beginning of the universe's explosion, when the expansion of the universe accelerated. With the separation of the two horizons, namely the value of $x$ is reduced, the entropic force between two horizon decreases, and when the $x=x_1$, the first time fellowship with the $x$ axis, the entropic force between the two horizon is zero, between the event horizon is not affected by external force, to maintain state of separation between two horizon, when the $x$ continue to reduce the negative entropic force between the two horizons, in the interval between two horizon deceleration separation. This case corresponds to our universe is slowing inflation. With the decrease of $x$, that is, the separation between the two horizons, the entropic force between the two horizons continues to decrease, and the entropic force curve gradually approaches the horizontal axis, and the entropic force between the two horizons approaches zero.

Comparing Fig.\ref{PVR3} with the curve of Lennard-Jones forces between two particles as they vary in location~\cite{Miao,Miao2}, we find that the curves obtained by completely different methods are so strikingly similar that the Lennard-Jones force of the two particles is intrinsically related to the entropic force between the two event horizons. Since the expansion of the universe is affected by various substances in the universe, the curve $F(x)-x$ in Fig.\ref{PVR3} reflects the influence of different parameters of space-time on the entropic force between the two horizons. If the entropic force is one of the internal forces driving the expansion of the universe, Fig.\ref{PVR3} shows that the accelerated expansion of our universe is influenced by various parameters. In particular, under the same parameters, the amplitude of the curve $F(x)-x$ is proportional to $\tilde{\alpha}$, indicating that the force between the two horizons is proportional to the GB factor, so the size of the GB factor $\tilde{\alpha}$ is proportional to the expansion rate of the universe.

\section{Discussion and Summary}
\label{con}
According to the discussion in the section (\ref{laten}), comparing the entropic force curve  $\tilde{F}(x)-x$ and $\tilde {F}_1 (x)-x$ of the location ratio between the two horizons in GB space-time given in Fig.\ref{PVR3} with the Lennard-Jones force changing with the location curve given in literature ~\cite{Miao,Miao2}, we find that the two curves obtained in different ways are very similar. The entropic force relation between the two horizons is derived from the theory within the framework of general relativity. It is derived from the theory of general relativity in combination with quantum mechanics and thermodynamics. The Lennard-Jones force between the two particles is based on the experiment. Although the method used is completely different, the results obtained by the two particles are surprisingly similar. This conclusion indicates that the entropic force between the two event horizons is related to the Lennard-Jones force between the two particles. In particular, the entropic force between the two horizons is proportional to GB factor $\tilde{\alpha}$, and the size of $\tilde{\alpha}$ plays a direct role in the acceleration between the two horizons. If entropic force is one of the kinetic energy driving the expansion of the universe, then the size of GB factor $\tilde{\alpha}$ is directly related to the acceleration of the expansion of the universe.

In the framework of general relativity, the entropic force of the interaction between the event horizon of a black hole and the event horizon of the universe deduced by theory has a very high similarity to the Lennard-Jones force between two particles verified by experiment. Therefore, the conclusion we give reveals the internal relationship between general relativity, quantum mechanics and thermodynamics. This analogy provides a new way for us to study the interaction between particles and the microscopic states of particles inside black hole, as well as the relationship between Lennard-Jones potential and the microscopic states of particles inside a normal thermodynamic system.

\section*{Acknowledgements}
We would like to thank Prof. Zong-Hong Zhu and Meng-Sen Ma for their indispensable discussions and comments. This work was supported by the Young Scientists Fund of the National Natural Science Foundation of China (Grant No.11205097), in part by the National Natural Science Foundation of China (Grant No.11475108), Supported by Program for the Natural Science Foundation of Shanxi Province, China(Grant No.201901D111315) and the Natural Science Foundation for Young Scientists of Shanxi Province,China (Grant No.201901D211441).

\end{document}